\documentclass[a4paper,12pt]{article}

\usepackage{amsmath}
\usepackage{hyperref}
\usepackage{cite}
\usepackage{amssymb}
\usepackage{braket}
\title{$Z_2$ Symmetry of $AdS_3 \times S^3 \times S^3 \times S^1$ Superstrings}
\author{Abbas Ali\footnote{Email:aali.ph@gmail.com} 
 and Mohsin Ilahi\\
		Physics Department,\\ Aligarh Muslim University,\\ Aligarh-202002, India }
\date{}
\begin{document}

\maketitle

\begin{abstract}
We know a lot about the superconformal field theory holographically dual to superstrings moving in $AdS_3 \times S^3 \times S^3 \times S^1$ background but the problem remains essentially unsolved in spite of having  been thoroughly examined from several different view points like the D-brane configurations, analysis of the spectra and the free field realizations. In this note we   solve a  more than two decades old problem encountered by Gukov, Martinec, Moore and Strominger in 2004 in the context of free field realization of large $N=4$ superconformal field theory. We do so by bringing into consideration insights from the  ADHM  instanton linear sigma models. In particular we use the $Z_2$ symmetry of the large $N=4$ superconformal algebra and the duality between its two small $N=4$ superconformal sub-algebras. This is the symmetry between two three spheres of above background. We use this symmetry to get a   complete free field realization of the large $N=4$ superconformal algebra. 
\end{abstract}

$AdS_5$ and $AdS_3$ are the two most intensely studied cases of the famous Maldacena Conjecture about duality between theories with gravity on one side and conformal field theories without gravity on other \cite{Maldacena:1997re, Witten:1998qj, Gubser:1998bc}. In the case of $AdS_3$ three distinct  cases have been investigated very extensively. These are the cases of super strings moving on manifolds with geometry $AdS_3 \times S^3 \times \mathcal{M}^4$ where $\mathcal{M}^4= T^4, K3~~\text{or}~~S^3 \times S^1$ and corresponding dual superconformal field theories.

According to Maldacena's original proposal the conformal field theory dual to strings moving on $AdS_3 \times S^3  \times T^4$ and $AdS_3 \times S^3  \times K3$ are the so called  symmetric orbifold CFTs $Sym^N(T^4)$ and $Sym^N(K3)$ respectively \cite{Giveon:1998ns}. The $AdS_3 /CFT_2$ case of the correspondence is formulated as follows. The 1+1 dimensional CFT describing the Higgs branch of the D1-D5 system ($Q_1$ branes of D1 type and $Q_5$  of D5 type aligned in a  specific manner) on $\mathcal{M}^4= T^4~\text{or}~K3$ is dual to Type IIB superstrings moving on the manifold $(AdS_3 \times S^3)_{Q_1Q_5} \times \mathcal{M}^4(Q)$. Here $Q$  specifies the dependence of $\mathcal{M}^4$ on  $Q_1$  and $Q_5$. 

The natural conclusion at this juncture is that the superconformal field theory dual to strings moving on $AdS_3 \times S^3\times S^3 \times S^1$ background should be $Sym(S^3 \times S^1)$ \cite{Elitzur:1998mm}. This idea runs into a number of problems. The problem was very thoroughly analyzed in Ref.\cite{Gukov:2004ym} from a number of different angles including the analysis of the supergravity solutions, structure of the moduli space, long strings, singular CFT, intersecting D-brane configurations,  free field realizations of the large $N=4$ superconformal symmetry, structure of the symmetric product orbifolds, an index for $Sym(S^3 \times S^1)$ and the comparison of the BPS and near BPS spectra on the two sides of the correspondence. The end product was that they ruled out all the existing proposals about the conformal field theory dual to strings on $AdS_3 \times S^3\times S^3 \times S^1$ geometry at that time.

After that the problem was taken up in Ref.\cite{Tong:2014yna}. The technique used there was of the  dynamics of the supersymmetric gauge theory living on the specifically designed  D-brane configuration.   Aspects related to the  D-brane configurations and gauge dynamics  for  $AdS_3 \times S^3\times K3$ and  $AdS_3 \times S^3\times T^4$ superstrings had been reviewed in Ref.\cite{David:2002wn} in  detail. The gauge theory considered in \cite{Tong:2014yna} for  $AdS_3 \times S^3\times S^3 \times S^1$ superstrings does give the required central charge but the brane configuration chosen was such that the geometry resulting from it does not go over to needed geometry, that is, $AdS_3 \times S^3\times S^3 \times S^1$. Thus the problem remained unsolved.

Next breakthrough came with the Ref.\cite{Eberhardt:2017fsi} whereby the proposal about $Sym(S^3 \times S^1)$ being the required CFT dual  was restored to its original place in Ref.\cite{Eberhardt:2017pty} by revisiting the spectrum analysis of Ref.\cite{deBoer:1999gea} (see also \cite{Chakraborty:2022iuk}). The correspondence was established by comparing the spectra as well as by matching the spectra on two sides when the D5 brane charge $Q_5^+$ is a factor of $Q_5^-$, where the charges $Q^\pm_5$ correspond to the sizes of the two $S^3$'s in $AdS_3 \times S^3\times S^3 \times S^1$. It is clear that this condition is generalizable to the condition that the two charges can be multiple of each other because two $S^3$'s are at equal footings. To settle the problem finally it is desirable to remove this limitation that $Q^\pm_5$ be multiple of each other.

In this note we track and trace the cause of this limitation and collect the data that, in our view, should help in the final resolution of the problem. We approach the problem  from the point of view of the free field realizations and believe that this angle takes us very close to the solution of the more than two  decades old problem - to get the complete free field realization of the large $N=4$ superconformal algebra. This should help us in solving the nearly three decade old problem to pinpoint the superconformal field theory holographically dual to  superstrings moving on $AdS_3 \times S^3\times S^3 \times S^1$ geometry. 

The fact that the charges of two different sets of D5 branes, $Q^\pm_5$, should be multiple of each other indicates that the some condition is already built up in the initial data that restricts the data to conditional values, not allowing $Q^\pm_5$ to vary independently. In Ref.\cite{Gukov:2004ym}  this realization is implicit. We shall further clarify it in a moment. 

We shall approach the problem from the point of view of the free field realizations of small, middle and large $N=4$ superconformal algebras.  The superconformal symmetry relevant for Type IIB superstrings  moving on $AdS_3 \times S^3  \times K3$ is the small N=4  \cite{Ademollo:1975an, Giveon:1998ns} while for Type IIB superstrings moving on $AdS_3 \times S^3  \times T^4$ it is the middle N=4 \cite{Ali:1993sd, Hasiewicz:1989vp, Ali:2000we, Ali:2003aa}. The superconformal symmetry relevant for the case of Type IIB superstrings moving on $AdS_3 \times S^3\times S^3 \times S^1$ is the large $N=4$\cite{Elitzur:1998mm, SEVRIN1988447, Ivanov:1988rt, Schoutens:1988ig, Spindel:1988sr, Goddard:1988wv, VanProeyen:1989me, Sevrin:1989ce, Ali:2000}. 

Superstrings on $AdS_3 \times S^3\times T^4$ are dual to the symmetric orbifold $Sym^n(T^4)=(T^4)^n/S_n$. But we do not know which background is described by the symmetric orbifold. In this regards Larsen and Martinec \cite{Larsen:1999uk} gave us some insights insights. Maldacena, Ooguri and Son constructed a solvable string theory on this background in terms of $SL(2, R)$ WZW theory \cite{Maldacena:2000hw, Maldacena:2000kv, Maldacena:2001km}. Yet we do not know to what precise dual it corresponds to. Existence of long strings in the spectrum \cite{Seiberg:1999xz} suggest that above superstrings with pure NS-NS flux can not be dual to symmetric orbifold. In worldsheet analysis this continuum vanishes in the continuum limit $k=1$. In this limit $l_{AdS}\sim l_{string}$ and the dual CFT becomes almost free. In this case the conserved currents correspond to massless higher spin fields in the bulk\cite{Fradkin:1987ks, Sundborg:2000wp, Klebanov:2002ja}. The symmetric orbifold contains a free subsector and we can suppose that it describes tensionless strings \cite{Gaberdiel:2014cha}. One conclusion is that the symmetric orbifold is located near the tensionless limit.

These issues were further investigated in Refs.\cite{Eberhardt:2018ouy, Eberhardt:2019qcl, Eberhardt:2019niq, Dei:2019osr}. In Ref. \cite{Eberhardt:2019ywk} a derivation of $AdS_3/CFT_2$ correspondence was presented. Correlation functions for the symmetric profuct were computed in Ref. \cite{Dei:2019iym}. Partition function for tensionless limit was computed in Ref. \cite{Eberhardt:2020bgq}. In Ref. \cite{Dei:2021yom} four pount functions for $AdS_3$ strings were computed. In Ref. \cite{Aharony:2024fid} discussed  the 1+1-dimensional superconformal field theory dual to type II string theory on $AdS_3 \times S^3\times T^4$ background, emphasizing the string theoretic aspects of AdS/CFT duality. In Ref. \cite{Eberhardt:2025sbi} problem of localization of $AdS_3$ sigma model was analyzed.

This is the lengthy background to the problem of finding the holographic dual CFT to superstrings moving on $AdS_3 \times S^3\times S^3\times S^1$ background. We shall focus on the free field realization of the dual superconformal field theory that has the large $N=4$ supersymmetry. 

In the rest of this note we shall do the following. First of all we shall collect the operators, corresponding operator  product expansions and the describe the features of the large $N=4$ superconformal algebra. Then we list the most common free field realization of this algebra and point out its short coming that in our view get carried over to the problem of finding the superconformal field theory dual to superstrings moving on $AdS_3 \times S^3\times S^3 \times S^1$ backgrounds. We then describe the procedure adopted by Gukov, Martinec, Moore and Strominger to overcome this short coming and point out why this route has lead to a road block. We argue that this is a red herring and that it should be possible to overcome it. Next we describe the procedure that overcomes the hurdle and present the resulting complete free field realization of $N=4$ superconformal algebra and discuss its implications.

The large $N=4$ superconformal algebra has sixteen generators: $T(z)$, $G^a(z)$, $A^{\pm i}(z)$,$Q^a(z)$ and $U(z)$, $a=1,2,3,4$ and $i=1,2,3$. Currents $A^{\pm i}(z)$ generate two $SU(2)$ Kac-Moody algebras and $U(z)$ a $U(1)$ algebra. Corresponding operator product expansions are given below:

\begin{eqnarray}
T(z)T(z') & = &\frac{c/2}{(z-z')^4}+\frac{2T(z')}{(z-z')^2}+\frac{\partial T(z')}{z-z'}+\cdots,\nonumber\\
T(z){\cal O}(z') & = &\frac{d_{\cal O}{\cal O}(z')}{(z-z')^2}+\frac{\partial {\cal O}(z')}{z-z'}+\cdots,\;{\rm where}\nonumber\\
{\cal O} &\in& \{G^a, A^{\pm i}, Q^a, U\}\; {\rm  with}\nonumber\\
d_{\cal O} &\in& \{3/2, 1, 1/2, 1\}\; {\rm respectively},\nonumber
\end{eqnarray}
    
\begin{eqnarray}
G^a(z)G^b(z') & = &\frac{2c/3\delta^{ab}}{(z-z')^3}+\frac{8[\gamma\alpha^{+i}_{ab}A^{+i}(z')+(1-\gamma)\alpha^{-i}_{ab}A^{-i}(z')]}{(z-z')^2}\nonumber\\
&+&\frac{\{2T(z')\delta^{ab}-4[\gamma\alpha^{+i}_{ab}\partial A^{+i}(z')+(1-\gamma)\alpha^{-i}_{ab}\partial A^{-i}(z')]\}}{z-z'}+\cdots,\nonumber\\
A^{+i}(z)G^a(z') & = &\alpha^{+i}_{ab}\left[\frac{G^b(z')}{z-z'}-\frac{2(1-\gamma)Q^b(z')}{(z-z')^2}\right]+\cdots,\nonumber\\
A^{-i}(z)G^a(z') & = &\alpha^{-i}_{ab}\left[\frac{G^b(z')}{z-z'}+\frac{2\gamma Q^b(z')}{(z-z')^2}\right]+\cdots,\nonumber\\
A^{\pm i}(z)A^{\pm j}(z') & = &-\frac{k^{\pm}/2\delta^{ij}}{(z-z')^2}-\frac{\epsilon^{ijk}A^{\pm k}(z')}{z-z'}+\cdots,\nonumber\\
A^{+i}(z)A^{-j}(z') & = & A^{\pm i}(z)U(z') = 0,\nonumber\\
Q^a(z)G^b(z') & = &\frac{2[\alpha^{+i}_{ab}A^{+i}(z')-\alpha^{-i}_{ab}A^{-i}(z')]}{z-z'}+\cdots,\nonumber\\
A^{\pm i}(z)Q^a(z') & = & \frac{\alpha^{\pm i}_{ab}Q^b(z')}{z-z'} +\cdots,\;\; U(z)Q^a(z')=0,\nonumber\\
U^a(z)G^a(z')&=&\frac{Q^a(z')}{z-z'},\; Q^a(z)Q^b(z') = -\frac{c}{12\gamma(1-\gamma)}\frac{\delta^{ab}}{z-z'}+\cdots,\nonumber\\
U(z)U(z')&=& -\frac{c}{12\gamma(1-\gamma)}\frac{1}{(z-z')^2}+\cdots.
\label{large1}   
\end{eqnarray}
Here 
\begin{equation}
    \alpha^{\pm i}_{ab}=\frac{1}{2}\epsilon_{iab} \pm \frac{1}{2}(\delta^i_a\delta^4_b-\delta^i_b\delta^4_a).
\end{equation}

Most of the conformal algebras have one independent parameter to label the corresponding representation, the central charge $c$. Even in case of conformal field theories based on Kac-Moody symmetries with Sugawara construction the level of the Kac-Moody algebra is related to the central charge and effectively we have only one parameter. In case of the large $N=4$ superconformal algebra above there are two independent parameters, the central charge $c$ and the parameter $\gamma$. We can also take $k^\pm$, the levels of the two internal Kac-Moody algebras as independent parameters. This counting of independent parameters is crucial to the problem we are focusing upon.

The relations between various parameters are given below.
\begin{eqnarray}
    c&=&\frac{6k^+k^-}{k^++k^-},\;\; \gamma=\frac{k^-}{k^++k^-},\;\; 1-\gamma=\frac{k^+}{k^++k^-},\nonumber\\
 k^+&=&   \frac{c}{6\gamma},\;\; k^-=\frac{c}{6(1-\gamma)}.
\end{eqnarray}

Under $k^+\leftrightarrow k^-$ or   $\gamma\leftrightarrow 1-\gamma$ or $SU(2)_+\leftrightarrow SU(2)_-$ where $SU(2)_+$ and $SU(2)_-$ are the two Kac-Moody subalgebras, we get another isomorphic $N=4$ superconformal algebra.

This corresponds to the $Z_2$ symmetry between the two three spheres of the $AdS_3 \times S^3\times S^3 \times S^1$ superstrings.

This duality is carried over to the two embeddings of the small $N=4$ superconformal algebra inside the large $N=4$ algebra.

This observation is the key to the solution to the problem of getting a complete free field realization of the large $N=4$ superconformal algebra starting with two Sevrin, Troost,  van Proeyen free field realizations. This will lead us to the solution of the problem that has remained unsolved since 2004.

Symmetries in real life occur not as abstract groups or algebras but as realizations in terms of concrete mathematical entities. In case of conformal algebras the free field realization provide an extremely powerful class of representations that have been put to great use. In case of the large $N=4$ superconformal algebra the most commonly used free field realization is the one that was given by  Sevrin, Troost and  van Proeyen in 1988. The generators of the large $N=4$ superconformal algebra are written in terms of eight `free' fields - $J^0(z), \psi^a(z)\; {\rm and}\; J^i(z)$. The explicit expressions are: 
\begin{eqnarray}
T(z)& = &-J^0J^0-\frac{1}{k+1}J^iJ^i+\psi^a\partial\psi^a,\nonumber\\
U(z)& = &\sqrt{k+1}J^0(z), ~~~~~Q^a(z)=\sqrt{k+1}\psi^a(z),\nonumber\\
 A^{+i}(z) &=& J^i+\alpha^{+i}_{ab}\psi^a\psi^b,\;\;A^{-i}(z) =\alpha^{-i}_{ab}\psi^a\psi^b,\nonumber\\
G^0(z)& = &2\left[J^0\psi^0+\frac{1}{\sqrt{k+1}}J^i\psi^i
+\frac{2}{\sqrt{k+1}}\psi^1\psi^2\psi^3\right],\nonumber\\
G^1(z)& = &2\left[J^0\psi^1
+\frac{1}{\sqrt{k+1}}(-J^1\psi^0+J^2\psi^3-J^3\psi^2)\right.\nonumber\\
&{}&\left.-\frac{2}{\sqrt{k+1}}\psi^0\psi^2\psi^3\right],\label{stv1}
\end{eqnarray}
with cyclic expressions for $G^2(z)$ and $G^3(z)$.
 
 It is simple to verify that the operators given in Eqn.(\ref{stv1})  generate the large $N = 4$ algebra with $k^+ = k+1$ and $k^-=1$. Here $k^+$ has a general value while $k^-$ has a fixed value. Thus while the superconformal algebra has two independent parameters, either $c$ and $\gamma$ or $k^\pm$ this free field realization has effectively only one parameter. This is the first indication that this free field realization is not the most general one possible. In fact this free field realization is merely a glorified small $N=4$ superalgebra. This can be seen as follows. This free field realization uses eight free fields : one $U(1)$ current $J^0(z)$, three $SU(2)$ currents $J^i(z)$, and four free fermions $\psi^a(z)$. Thus there are only eight `free' fields which is the number of generators of the small $N=4$ superconformal algebra. The number of generators in the large $N=4$ superconformal algebra is sixteen. Thus there is a fifty percent deficit in the number of free fields as compared to the number of generators. This is another indication that this free field realization has a limitation. In fact we see that the level of the SU(2) Kac-Moody algebra generated by the currents $A^{-i}(z)$ has a fixed value equal to one   because the corresponding currents are realized as fermion bilinears and there is no genuine SU(2) current sitting in these unlike the SU(2) currents $A^{+i}(z)$ where we do have a genuine SU(2) current, $J^i(z)$, sitting in.  In our view this limitation gets carried all the way up to the problem of finding the superconformal field theory dual to superstrings moving on $AdS_3 \times S^3\times S^3 \times S^1$ geometry. 
 
 There is another manner in which this limitation manifests itself. We can make two independent In\"on\"u-Wigner contractions of the large $N=4$ superconformal algebra. This will result in the middle $N=4$ superconformal algebras in two independent ways. On the other hand the the Sevrin, Troost and van Proeyen's free field realization yields only one contraction because it has only one free parameter $k$. This is because in the contraction we have to take singular limit of the level of an $SU(2)$ Kac-Moody subalgebra and in case of above free field realization only one level has a general value and we can take its singular limit. The level of the other $SU(2)$ Kac-Moody subalgebra is fixed and we can not take its singular limit and hence no contraction is possible.

To see the two independent contractions of the large $N=4$ superconformal algebra and only only contraction of the Sevrin, Troost, van Proeyen's free field realization we refer the reader to Ref. \cite{Ali:2003aa}.

Thus the complete free field realization of large N=4 superconformal symmetry should be such that we can take two In\"on\"u-Wigner contractions and in each case still end up in general free field realizations of the middle $N=4$ superconformal algebra. In other words the Sevrin-Troost-van-Proeyen realization should be augmented in such a way so as to permit two In\"on\"u-Wigner contractions.

One way to overcome this limitation is to do tensoring of two copies of the Sevrin, Troost and van Proeyen's free field realization. Each copy will contribute one general parameter and in the end product we shall get the requisite number, two, of parameters. 

In Ref.\cite{Gukov:2004ym} a slightly different route was adopted. They assumed that a complete free field realization of the large $N=4$ superconformal algebra. Then they took two copies of this free field realization. The $SU(2)$ levels of the two copies  were called $k_1^\pm$ and  $k_2^\pm$ respectively. The energy momentum tensor of the two copies were called $T_1(z)$ and $T_2(z)$ respectively. The $U(1)$ currents of the two copies were called $U_1(z)$ and $U_2(z)$ respectively.

The proposal in Ref.\cite{Gukov:2004ym} had the following prescription:
\begin{eqnarray}\label{gmmsprop}
T(z)&=&T_1(z)+T_2(z)+ \frac{1}{2}\partial(aU_1(z)+bU_2(z))\ \nonumber \\
G^a(z)&=&G^a_1(z)+G^a_2(z)+\partial(aQ^a_1(z)+bQ^a_2(z))  \nonumber \\
A^{\pm i}(z)&=&A_1^{\pm i}(z)+A_2^{\pm i}(z),  ~U(z) = U_1(z)+U_2(z),\nonumber \\
Q^a(z)&=&Q^a_1(z)+Q^a_2(z).
\end{eqnarray}

Here $A^{\pm i}_1(z)$ and $A^{\pm i}_2(z)$ as well as $Q^a_1(z)$ and $Q^a_2(z)$ notations are obvious. By the closure of the large $N=4$ superconformal algebra the values of the constants $a$ and $b$ were determined to be 
\begin{eqnarray}\label{consab}
a = 2\frac{k_1^+k_2^--k_1^-k_2^+}{k_1(k_1+k_2)},
~b = 2\frac{k_2^+k_1^--k_2^-k_1^+}{k_2(k_1+k_2)},
\end{eqnarray}
where $k_1=k_1^++k_1^-$ and $k_2=k_2^++k_2^-$.

Using the prescription $U(z)=\sqrt{\frac{k}{2}}\phi(z)$ for the $U(1)$ currents we get two orthogonal combinations
\begin{equation*} \label{orthocomb1}
U_+(z)=\sqrt{\frac{k_1}{k_1+k_2}}\phi_1(z) +\sqrt{\frac{k_2}{k_1+k_2}}\phi_2(z) 
\end{equation*}
and
\begin{equation} \label{orthocomb2}
U_-(z)=\sqrt{\frac{k_2}{k_1+k_2}}\phi_1(z) -\sqrt{\frac{k_1}{k_1+k_2}}\phi_2(z) 
\end{equation}
and for the stress tensor we get
\begin{equation} \label{stressten}
T(z)=-\frac{1}{2}(\partial\phi_-(z))^2 +\frac{1}{2}Q_{12}(\partial)^2\phi_-(z)
\end{equation}
with
\begin{equation}
    U_\pm(z)=\partial_\pm\phi(z),
\end{equation}
and 
\begin{equation}\label{q12}
Q_{12}=\sqrt{2}\frac{k_1^+k_2^--k_2^+k_1^-}{\sqrt{k_1k_2(k_1+k_2)}}. 
\end{equation}
From this we get
\begin{equation}\label{q123}
Q_{(12)3} \neq Q_{(1(23)}.
\end{equation}
This means that the tensoring procedure used above is not associative. Thus  the search for a complete free field realization of the large $N=4$ superconformal symmetry by tensoring two copies of complete free field realizations was abandoned. This shut the door of progress on this front for two decades. 

We shall not investigate the causes of the non-associativity encountered above in the present note. Our main push in this paper is to demonstrate that this hurdle is a red herring. We shall also point out that there is a way out of this impasse.

To overcome the non-associativity we shall make use of that mathematical structure where associativity is built in - the group structure. In fact we shall take help from one of the simplest symmetry groups - $Z_2$. For this we have to collect some information about the already discovered symmetry structure of $AdS_3$ superstrings.  We were lead to this structure from the analysis of the symmetry structure of ADHM instanton sigma models\cite{Witten:1994tz, Ali:2023csc, Ali:2023ams, Ali:2023icn, Ali:2023xov, Ali:2024amc, Ali:2025ntc, Ali:2025jcu}.

Original ADHM instanton sigma model was constructed by Witten in Ref.\cite{Witten:1994tz}. In this construction the conditions to have $N=4$ supersymmetry turn out to be the same as the ADHM conditions. It is widely believed that such theories flow to superconformal field theories in the infrared. We know that there are three different $N=4$ superconformal symmetries in two dimensions: small, middle and large. We believe that Witten's ADHM instanton sigma model will flow in the infrared limit to a field theory with small $N=4$ superconformal symmetry. The $N=4$ supersymmetry in Witten's construction uses an $SU(2)$ symmetry $F'$. Apart from $F'$ there is another $S(2)$ symmetry in the initial field content which is termed $F$. Witten had suggested that an alternative model using this should be constructed. The resulting model is called the complementary ADHM instanton sigma model. It was constructed in Ref. \cite{Ali:2023csc}. We believe that this too flows in the infrared to a small $N=4$ superconformal field theory. There is a duality between the complementary and the original models. This is the duality between the two $S(2)$ symmetries $F$ and $F'$.

Witten also suggested a construction of a model that uses both of the $SU(2)$ symmetries $F$ and $F'$. We did corresponding construction in Ref.\cite{Ali:2023icn}.  Since this model has two $SU(2)$ symmetries it should flow in the infrared to a large $N=4$ superconformal field theory. We believe that the ADHM instanton sigma models can be mapped onto $AdS_3$ superstrings. The original and the complementary ADHM instanton sigma models flow in the infrared to such small $N=4$ superconformal symmetries that are relevant for superstrings moving in $AdS_3\times S^3\times K3$ backgrounds. In Ref.\cite{Ali:2023xov} (see also \cite{Ali:2024amc}) we realized that there should be two such theories related to each other by duality. The complete ADHM instanton sigma model of Ref.\cite{Ali:2023icn}, on the other hand, should flow to a large $N=4$ superconformal theory that is relevant for superstrings moving on  $AdS_3\times S^3\times S^3\times S^1$ backgrounds. The duality $F$ and $F'$ of the original and complementary models now becomes a $Z_2$ symmetry between them. This is the $Z_2$ symmetry between the two $S^3$'s of the background.

Witten also suggested a construction of a model that uses both of the $SU(2)$ symmetries $F$ and $F'$. We did corresponding construction in Ref.\cite{Ali:2023icn}.  Since this model has two $SU(2)$ symmetries it should flow in the infrared to a large $N=4$ superconformal field theory. We believe that the ADHM instanton sigma models can be mapped onto $AdS_3$ superstrings. The original and the complementary ADHM instanton sigma models flow in the infrared to such small $N=4$ superconformal symmetries that are relevant for superstrings moving in $AdS_3\times S^3\times K3$ backgrounds. According to the analysis of Ref.\cite{Ali:2023xov}   there should be two such theories related to each other by duality. The complete ADHM instanton sigma model of Ref.\cite{Ali:2023icn}, on the other hand, should flow in infrared to a large $N=4$ superconformal theory that is relevant for superstrings moving on  $AdS_3\times S^3\times S^3\times S^1$ backgrounds. The duality $F$ and $F'$ of the original and complementary models now becomes a $Z_2$ symmetry in this case. This is the $Z_2$ symmetry between the two $S^3$'s of the background. These aspects have been organized in detail in Ref. \cite{Ali:2024amc}

We can also answer the question as to where is the middle $N=4$ superconformal symmetry in this rich structure. It is obtained by In\"on\"u-Wigner contraction of the large $N=4$ superconformal algebra \cite{Ali:1993sd, Hasiewicz:1989vp, Ali:2000we, Ali:2003aa}. From the insights that we gained from the structure of the ADHM instanton sigma models we realize that this contraction can be done in two ways and we get two middle $N=4$ superconformal algebras related to each other by a duality. In case of $AdS_3$ superstrings we get $AdS_3\times S_+^3\times T^4$ and $AdS_3\times S_-^3\times T^4$ by two different Penrose limits  of $AdS_3\times S_+^3\times S_-^3\times S^1$ of background.

Our next task is to make use of these insights about the symmetry structure of ADHM instanton sigma models,  $AdS_3$ superstrings and $N=$ superconformal algebras to reach a complete free field realization of the large $N=4$ superconformal algebra.

The $Z_2$ symmetry must be built into the free field realization. To this end we begin by introducing two genuine $SU(2)$ currents among the {\it free} fields - $J^\pm(z)$ and define
\begin{equation}
A^{\pm i}(z)= J^{\pm i}(z) \mp \psi^0(z)\psi^i(z) \mp \chi^0(z)\chi^i(z)+ \epsilon^{ijk}[\psi^j(z)\psi^k(z)+\chi^j(z)\chi^k(z)].\\
\label{2su2}
\end{equation}
Here the {\it free} field $SU(2)$s of Sevrin, Troost and van Proeyen realization has been labeled as $J^{+i}(z)$ and $J^{-i}(z)$ are the new {\it free} $SU(2)$ currents. To maintain supersymmetry we have also introduced corresponding fermions $\chi^a(z)$.

This in turn requires introduction of second $U(1)$ current
\begin{equation}
K^0(z)=\partial\varphi(z)\label{su1}   
\end{equation}
in addition to the one already present in the Sevrin, Troost and van Proeyen realization, that is,
\begin{equation}
J^0(z)=\partial\phi(z).\label{fu1}   
\end{equation}

The fact that the large $N=4$ superconformal algebra has only one $U(1)$ is solved by putting different background charges on both $\phi(z)$ and $\varphi(z)$ such that one linear combination is a genuine $U(1)$ while other one has a background charge.

This is the point where the $Z_2$ symmetry plays the crucial role.

The expression for the genuine $U(1)$ is
\begin{equation}
    U(z) = \sqrt{\frac{c}{6\gamma}}J^0(z)+ 
\sqrt{\frac{c}{6(1-\gamma)}}K^0(z).\label{u1}
\end{equation}
The $Z_2$ symmetry here is $\phi\leftrightarrow\varphi$, $\gamma\leftrightarrow 1-\gamma$. The coefficients of $J^0(z)$ and $K^0(z)$ in Eqn. (\ref{u1}) have been chosen to give the correct normalization of the genuine $U(1)$ current $U(z)$ in accordance with the algebra in Eqn.(\ref{large1}).

Same process also gives us the genuine spin 1/2 currents $Q^a$'s as
\begin{equation}
Q^a(z) = \sqrt{\frac{c}{6\gamma}}\psi^a(z)+ \sqrt{\frac{c}{6(1-\gamma)}}\chi^a(z).\label{q}
\end{equation}
The coefficients of $\psi^a$ and $\chi^a$ are once again chosen to give correct normalizationn of $Q^a(z)$ according to the OPEs in Eqn.(\ref{large1}). The $Z_2$ here is  $\psi\leftrightarrow\chi$ and $\gamma\leftrightarrow 1-\gamma$.

We still need expressions for five operators, $T(z)$ and $G^a(z)$, in terms of free fields. Conformal spins and weights dictate the following expression for $T(z)$:
\begin{eqnarray}
T(z) = &-&J^0J^0+ \sqrt{\frac{\gamma}{6c}}(6-c)\partial J^0 - K^0K^0 -\sqrt{\frac{1-\gamma}{6c}}(6-c)\partial K^0 \nonumber \\
&-&\frac{6\gamma}{c}J^{+i}J^{+i} - \frac{6(1-\gamma)}{c}J^{-i}J^{-i}+\psi^a \partial \psi^a+\chi^a \partial \chi^a.\label{t} \\
\end{eqnarray}
Here the coefficients of $\partial J^0$ and $\partial K^0$ are chosen to make $U(z)$ a genuine $U(1)$ current and to give the correct central charge. There is no  $Z_2$ symmetry in this case under  $\psi\leftrightarrow\chi$ and $\gamma\leftrightarrow 1-\gamma$,
$J^0\leftrightarrow K^0$, $J^{+i}\leftrightarrow J^{-i}$. Instead we get the energy momentum tensor for an isomorphic large $N=4$ superconformal algebra.

Finally we must have the free field realization of the supercurrents $G^a(z)$. These have the following expression 
\begin{eqnarray}
G^a(z) = &2& [J^0\psi^a+\sqrt{\frac{\gamma}{6c}}(6-c)\partial\psi^a
+K^0\chi^a-\sqrt{\frac{(1-\gamma)}{6c}}(6-c)\partial\chi^a
\nonumber\\
&+& \sqrt{\frac{6\gamma}{c}}(\alpha^{+i}_{ab}J^{+i}\psi^b+2\epsilon_{abcd}\psi^b\psi^c\psi^d)\nonumber\\
&+&\sqrt{\frac{6(1-\gamma)}{c}}(\alpha^{-i}_{ab}J^{-i}\chi^b+2\epsilon_{abcd}\chi^b\chi^c\chi^d)].
\label{g}\nonumber\\
\end{eqnarray}
Once again there is no  $Z_2$ symmetry here too under  $\psi\leftrightarrow\chi$ and $\gamma\leftrightarrow 1-\gamma$,
$J^0\leftrightarrow K^0$, $J^{+i}\leftrightarrow J^{-i}$ and instead we get supercurrents for the isomorphic large $N=4$ superconformal algebra mentioned above.

The expressions in Eqs. (\ref{u1}), (\ref{q}), (\ref{t}) and (\ref{g}) give us a complete free field realization of the large $N=4$ superconformal algebra.

This gives another automorphic free field realization of the large $N=4$ superconformal algebra under  $ \gamma \leftrightarrow 1-\gamma$.  Its implications are discussed next.

The free field realization above is complete in the sense that we can now get two different singular  limit of the same realization and get $AdS_3 \times S_{+}^3 \times T^4$ and     $AdS_3 \times S_{-}^3 \times T^4$  geometries with general radii for $S_{\pm}^3$ because remaining $SU(2)$'s have general values of the levels.  

On the geometrical side we learn that the $AdS_ 3 \times S^3 \times S^3 \times S^1$ geometry should go to $AdS_3 \times S_{}^3 \times T^4$ geometry in the Penrose limit in case of  geometry and In\"on\"u-Wigner  contraction of free field realization.  Moreover this should happen in two independent ways to give $AdS_3 \times S_{+}^3 \times T^4$ and     $AdS_3 \times S_{-}^3 \times T^4$  geometries.

Now let us take stock of the situation and see what has been achieved and what directions open up for future progress.

In summary we have removed the main stumbling block in one of the many approaches, namely the free field realizations of the relevant superconformal algebra, applied to find the superconformal field theory dual to superstrings moving on $AdS_ 3 \times S^3 \times S^3 \times S^1$ geometry. We have listed the relevant free field realization of the corresponding superconformal algebra. This, in our view, opens the door for solution to the main problem that is by now a quarter of a century old and has eluded in spite of very robust investigations by several groups. 

At the moment we do not know the route to find the superconformal field theory that is holographically dual to superstrings moving on the $AdS_ 3 \times S^3 \times S^3 \times S^1$ background geometry using the complete free realization of the large $N=4$ superconformal algebra mentioned above but we do believe that this is one of the key ingredient to lead us to a complete and final solution to this problem. We also believe that the insights obtained by the corresponding analysis will help us in finding solutions to every problem that is related to the problem under consideration. 

There is a large number of aspects upon which the resulting solution has a bearing. For example the D-brane configuration that would give above geometry in the decoupling limit is not known. Insights from the present note should help in corresponding investigations. There is another quarter of a century old problem that needs solution. It is the problem of the singularity of the field theory associated with the D1/D5 system \cite{Seiberg:1999xz} because of latter's close connection with $AdS_3$ superstrings. We believe our investigations in the present note would be helpful in thinking about this problem too. It would be interesting to work out the highest weight representations of the large $N=4$ superconformal algebra in the light of the complete free field realization given above and find out what additional insights. There are many other related problems whose investigation would be helped by the results of the present note. Of course the highest priority issue is to get to the superconformal field theory  that is holographically dual to superstrings moving on $AdS_ 3 \times S^3 \times S^3 \times S^1$ backgrounds by taking advantage of the complete free field realization  above.

In related developments we mention the following. Integrability of superstrings on $AdS_3$ backgrounds has been discussed, for example, in Refs. \cite{Chakraborty:2022iuk, Sundin:2012gc, Dei:2018jyj} and the 
BMN Dynamics in Ref. \cite{Rughoonauth:2012qd}. In Ref.\cite{Papadopoulos:2024uvi} Papdopoulos and Witten gave a direct proof of the fact that in two dimensions scale invariance implies conformal invariance. Superstrings on $AdS_3$ backgrounds are closely related to stringy generalization of BTZ black holes \cite{Ali:1992mj}. It would be interesting to work out the implications of the present investigations in this context.

\textit{Acknowledgments}: This work was carried out as part of Mohsin Ilahi's Ph.D. thesis. AA thanks Professor V Ravindran and the Institute of Mathematical Science, Chennai where the task of revision of this note was done.

\end{document}